\documentclass[10pt,twocolumn]{article}
\usepackage[letterpaper,margin=0.72in,columnsep=0.24in]{geometry}
\usepackage{amsmath,amssymb}
\usepackage{booktabs}
\usepackage{graphicx}
\usepackage[T1]{fontenc}
\usepackage{microtype}
\usepackage{xcolor}
\usepackage{enumitem}
\usepackage{multirow}
\usepackage{float}
\usepackage{placeins}
\usepackage{url}
\usepackage[numbers,sort&compress]{natbib}
\usepackage[colorlinks=true,allcolors=blue!55!black]{hyperref}

\setlist[itemize]{leftmargin=*,nosep}
\setlist[enumerate]{leftmargin=*,nosep}
\title{olBlend: Language-Guided Molecular Blend Retrieval\\
with Mixture Odor Prediction}
\author{%
Dian Jin\textsuperscript{1,2}\thanks{Equal contribution.},
Zhechang Chen\textsuperscript{1,2}\footnotemark[1],
Qilin Liang\textsuperscript{1,2},
Yonghui Luo\textsuperscript{1,2}\\
Zixun Fu\textsuperscript{3},
Defeng Sun\textsuperscript{3},
Yancheng Yuan\textsuperscript{3},
Xiaoming Tao\textsuperscript{1,2}\\[0.4em]
\small\textsuperscript{1}Research Institute for Intelligent Wearable Systems,
The Hong Kong Polytechnic University\\
\small\textsuperscript{2}School of Fashion and Textiles,
The Hong Kong Polytechnic University\\
\small\textsuperscript{3}Department of Applied Mathematics,
The Hong Kong Polytechnic University}
\date{}

\begin{document}
\maketitle

\begin{abstract}
Converting open-ended natural-language odor descriptions into molecular mixtures that can be physically prepared from an available inventory remains an important challenge in computational olfaction. We propose olBlend, a language-guided framework for molecular mixture retrieval. olBlend first uses MixPOM to predict odor probabilities for candidate molecular pairs and stores these predictions for retrieval. It then uses a language model to translate each odor description into three types of semantic requirements: lead attributes that define the main odor, ambient attributes that describe its background, and absent attributes that should be avoided. A structured scoring algorithm uses the precomputed odor probabilities to score how well each candidate mixture meets these requirements, then retrieves and ranks mixtures from the available inventory. Evaluation on established odor datasets shows that MixPOM improves mixture odor prediction over averaged single-molecule predictions. Retrieval comparisons across natural-language queries demonstrate how background and exclusion requirements guide mixture selection. We further prepare retrieved formulations as physical odor samples, demonstrating the workflow from natural-language descriptions to realizable molecular formulations.
\end{abstract}

\section{Introduction}
Natural language is increasingly becoming an important way for people to generate perceptual content. In recent years, advances in multimodal AI have enabled users to generate desired images \cite{ramesh2021zero,peebles2023scalable}, sounds \cite{liu2023audioldm,copet2023simple}, and even 3D scenes \cite{yi2024gaussiandreamer,hu2025turbo3d} from open-ended natural language descriptions. However, extending this paradigm to the olfactory domain remains challenging. For example, given a description such as “freshly baked strawberry cake,” few existing studies can effectively interpret the target odor characteristics implied by the description and determine which odor molecules, and in what combinations, can produce a corresponding olfactory experience in the physical world.

This difficulty arises from both the intrinsic characteristics of olfactory perception and the physical constraints of practical odor generation systems. In vision, images can be represented as spatial arrangements of RGB color components, while sounds can be decomposed into frequency components. In contrast, human olfaction still lacks a widely accepted, low-dimensional set of perceptual primitives \cite{fulton2024common}. More fundamentally, even the cardinality of human olfactory perceptual space remains unresolved, with estimates of the number of discriminable odors varying by many orders of magnitude \cite{bushdid2014humans,meister2015dimensionality,gerkin2015number}.

In recent years, machine learning has significantly improved the ability to predict perceptual odor properties from molecular structures. Existing methods can learn mappings from individual molecules  \cite{keller2017predicting,lee2023principal} or molecular mixtures \cite{snitz2013predicting,sisson2025deep,kang2026aromma} to odor descriptors. These advances provide powerful models of the forward relationship between molecular composition and odor perception. However, odor generation requires solving the problem in the opposite direction: given a desired perceptual target, one must determine which molecules should be combined to produce the corresponding olfactory experience.

This inverse problem becomes particularly challenging in practical odor-generation systems. Such systems can typically accommodate only a finite set of odorants, constrained by hardware capacity and the physicochemical properties of the odor sources \cite{prasetyawan2024odor,wen2019development}. Consequently, the goal is not to search freely over the space of possible molecules, but to identify a mixture from a predefined inventory that is physically realizable and perceptually matches an open-ended natural-language description as closely as possible. Despite progress in molecular odor prediction, a systematic solution to this language-guided, inventory-constrained mixture retrieval problem is still lacking.

To address this problem, we propose olBlend, a natural-language-driven odor realization system based on mixture odor prediction under molecular inventory constraints. Given a set of molecules available in a laboratory or an odor generation device, olBlend first uses a trained mixture odor prediction model to estimate the odor properties of candidate molecular combinations, and then builds an odor database for the current inventory. When a user provides a natural-language odor description, the system uses a large language model to construct three mutually exclusive semantic sets from a predefined vocabulary: {lead}, which defines the main identity of the target odor; {ambient}, which provides supporting background characteristics; and {absent}, which contains odor properties that clearly conflict with the user's intention. During online ranking, olBlend combines the mean log-probabilities of lead and ambient presence and explicit-conflict absence, together with a co-activation-weighted penalty on unrequested descriptor probabilities. It then retrieves the realizable molecular mixtures that best match the user's semantic target from the current inventory.

We demonstrate language-guided formulation retrieval over 4,005 candidates from an inventory of 89 odorants and prepare five retrieved formulations as physical samples. Comparisons across open-ended queries show how the structured objective incorporates background and exclusion requirements into candidate selection, while prediction benchmarks and ablations demonstrate the value of learning mixture-specific corrections to single-molecule predictions. The proposed framework converts forward mixture odor prediction into an inverse retrieval system driven by user requirements, providing a practical path from natural-language descriptions to physically preparable formulations. It can also support applications such as formulation exploration for fragrance experts, personalized scent design, automatic odor generation in virtual environments, and digital odor search.

The main contributions of this work are summarized as follows:

\begin{enumerate}

    \item We propose olBlend, a natural-language-driven molecular mixture retrieval framework. olBlend parses a natural-language odor description into three types of attributes: lead, ambient, and absent. Given the available molecular inventory, it combines a role-normalized Bernoulli log-likelihood for desired and absent attributes with a co-activation-weighted regularizer on additional odor properties to retrieve physically realizable mixtures that best match the target description.

    \item We develop POM+ and MixPOM to predict candidate odor profiles. POM+ provides a single-molecule prior, and MixPOM learns mixture-specific corrections that capture non-additive interactions between components. This design combines single-molecule odor knowledge with mixture supervision to guide formulation retrieval.

    \item We demonstrate language-guided retrieval across five open-ended queries over 4,005 candidates from an inventory of 89 odorants, evaluate POM+ and MixPOM through prediction benchmarks and ablations, and physically prepare five retrieved formulations. Together, these experiments establish the predictive basis of the framework and demonstrate its workflow from language to physical samples.

\end{enumerate}

\section{Related Work}

\subsection{Molecular and Mixture Odor Prediction}

Predicting human odor perception from molecular structure is a central problem in computational olfaction. Early quantitative structure--odor relationship methods mainly relied on hand-crafted molecular descriptors, while recent studies increasingly use graph neural networks to learn odor-relevant representations directly from molecular structures. Lee et al. introduced the Principal Odor Map (POM), which learns a mapping from molecular graphs to perceptual odor descriptors using a graph neural network. Trained on the GS-LF dataset, POM provides a perceptually meaningful representation space for single molecules and establishes a strong basis for data-driven molecular odor prediction \cite{lee2023principal}.

Real-world odors, however, are usually mixtures of multiple molecules, and their perceptual properties cannot in general be obtained by simply combining the odors of individual components. Experimental studies have reported suppression, antagonism, and other nonlinear effects in odor mixtures \cite{zak2020antagonistic}. Recent machine learning methods have therefore started to extend structure--odor prediction from single molecules to mixtures. Sisson et al. constructed the Blended Pair (BP) dataset and trained graph neural networks to directly predict odor descriptors for pairs of aroma chemicals \cite{sisson2025deep}. Their analysis also showed that the odor labels and learned representations of a blend cannot be well explained by simple combinations of its component odors, further indicating the importance of non-additive mixture effects.

Another line of work focuses on learning perceptual representations of mixtures. Tom et al. proposed POMMix, in which the CheMix module applies molecule-wise self-attention and permutation-invariant aggregation to POM embeddings to obtain mixture representations \cite{tom2025does}. POMMix successfully models perceptual similarity between molecular mixtures and shows that attention-based aggregation is effective for capturing relations among mixture components. More recently, AROMMA jointly models single molecules and molecular pairs in a shared olfactory space. It combines a large-scale pretrained molecular encoder with an attention-based mixture aggregator, knowledge distillation, and class-aware pseudo-labeling to address the sparse and heterogeneous annotations of single-molecule and mixture datasets \cite{kang2026aromma}. PhysSim further explores a physics-inspired latent interaction model for predicting perceptual similarity between molecular mixtures \cite{kim2026physics}.

Despite this progress, an important gap remains in how non-additive mixture effects are represented in descriptor-level odor prediction. Existing methods either predict mixture similarity, as in POMMix and PhysSim, or learn mixture descriptors directly from an aggregated mixture representation, as in Sisson et al. and AROMMA. In these models, the contribution already explained by the individual molecules and the perceptual change introduced by mixing are not explicitly separated. As a result, the model must learn both the basic odor properties of the components and their interaction effects within the same mixture prediction function, which is particularly difficult when mixture annotations are sparse.

\subsection{Odor Generation and Retrieval}

Compared with the forward problem of predicting odor perception from molecules, the inverse problem of finding a physically realizable molecule or mixture for a desired perceptual target has received less attention. Early work on odor reproduction mainly aims to approximate a target scent using a small set of predefined odor components. These methods typically start from physical measurements or a predefined odor representation and then optimize the mixture ratios of available components. For example, Iseki et al. studied odor reproduction using a multi-component olfactory display, showing that mixtures of a fixed set of odor components can approximate selected target odors \cite{iseki2022study}. Prasetyawan and Nakamoto further reviewed and developed odor reproduction methods based on a small set of odor components \cite{prasetyawan2024odor}.

More recent work has introduced machine learning into this inverse process. Aleixandre et al. combined odor descriptor prediction from mass spectra, odor-component decomposition, and gradient-based optimization to generate mixtures matching a specified set of odor descriptors \cite{aleixandre2024automatic}. Their system demonstrates that a learned odor predictor can be used in the reverse direction to guide the creation of physical scent mixtures. They later introduced a generative diffusion model for creating scents conditioned on required odor descriptors \cite{aleixandre2025generative}. These studies provide important steps toward computational scent creation. However, their targets are specified in a predefined descriptor space rather than as open-ended natural-language descriptions. The generated recipes are also tied to a predefined set of odor components or essential oils.

Recent studies have started to use language models to connect natural language with olfactory systems. Sniff AI investigates whether language-model embeddings can identify scents from human natural-language descriptions \cite{zhong2024sniff}. Its main goal is perceptual alignment and scent identification rather than the construction of new molecular mixtures. AromaGen goes one step further and directly maps free-form text, images, or speech to mixture ratios over 12 predefined base odorants using the latent olfactory knowledge of a multimodal language model \cite{wen2026aromagen}. It also allows users to iteratively modify the generated aroma through natural-language feedback. More recently, a manufacturable text-to-scent framework uses an LLM to select fragrance notes from a curated palette and a deterministic rule engine to enforce top--middle--base structure and valid mixture weights \cite{kim7051853manufacturable}. These systems demonstrate the potential of natural language as an interface for odor generation.

However, a gap remains between semantic scent generation and molecular-level odor prediction. Existing language-driven approaches mainly rely on the olfactory knowledge of an LLM to directly select odorants or fragrance notes from a fixed palette. As a result, the generated recipe is not explicitly evaluated by an odor model trained on molecular and mixture perception data. In particular, molecular interactions that can change the perceived odor after mixing are not considered during semantic-to-recipe mapping. Moreover, existing approaches mainly focus on generating desired properties, while undesired or unrequested odor attributes are rarely modeled explicitly.

\section{Method}
\label{sec:method}

\begin{figure*}[ht]
    \centering
    \includegraphics[page=1,width=0.98\textwidth]{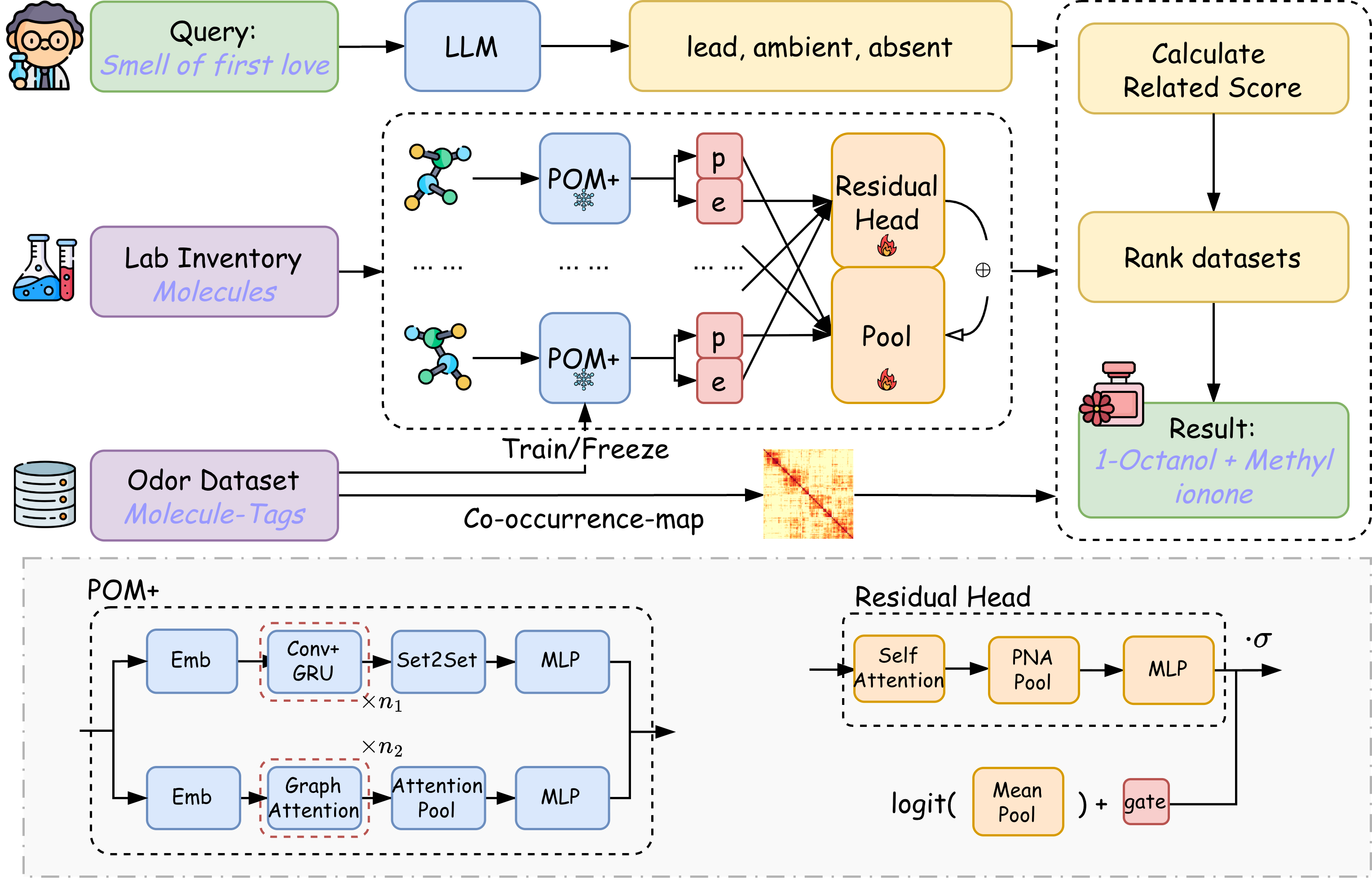}
    \caption{Overview of olBlend.}
    \label{fig:overview}
\end{figure*}

The architecture of our proposed method olBlend is shown in Figure~\ref{fig:overview}.
We consider single-molecule candidates and binary mixtures at fixed concentration ratios.

\subsection{Problem Setup}

Given a natural-language odor query $q$ and an available molecular inventory $\mathcal I$, olBlend searches the candidate set $\mathcal C=\{S\subseteq\mathcal I\}$ and returns the highest-scoring single molecule or blender pair:
\begin{equation}
\label{eq:retrieval}
S^*(q)=\arg\max_{S\in\mathcal C}r(S,q).
\end{equation}
Here, $r$ is a function that measures compatibility between the query and a candidate's predicted odor profile. The output specifies component identities for production.

\subsection{Predicting Candidate Odor Profiles}
\label{sec:predictor}

Mixture-level supervision is limited, whereas single-molecule annotations already provide substantial information about the odor attributes of individual components. Recent psychophysical evidence shows that a linear average of component odor profiles predicts mixture profiles well across the mixtures tested, supporting the use of component predictions as a prior \cite{pellegrino2026odors}. Suppression, masking, synergy, and other nonlinear interactions may nevertheless cause departures from this reference \cite{zak2020antagonistic}. We therefore use single-molecule predictions as a strong prior and let the mixture model learn corrections from mixture-level supervision.

To construct this prior, our proposed POM+ combines a POM \cite{lee2023principal} branch with a GraphGPS \cite{rampavsek2022recipe} branch to incorporate local message passing and global attention. For molecule $m_i$, we fuse the two branches' outputs  to obtain $\mathbf p_i$ and $\mathbf e_i$. The former provides the single-molecule odor reference, and the latter serves as input to MixPOM. Both branches remain frozen during mixture training. The predictive benefit of POM+ is evaluated in Section~\ref{sec:prediction_results}.

Single-molecule and Pair descriptors are aligned to a shared vocabulary, with the aligned single-molecule prediction denoted by $\widetilde{\mathbf p}_i$. 
Vocabulary alignment and model settings are detailed in Appendix.


For a component set $S$, MixPOM uses the average component prediction as a reference and a mixture encoder to learn relationships among the components:

\begin{equation}
\label{eq:mixture_repr}
\begin{aligned}
\bar{\mathbf p}(S)&=\frac{1}{|S|}\sum_{m_i\in S}\widetilde{\mathbf p}_i,\\
\mathbf h_S&=\operatorname{Pool}\!\left(\operatorname{Enc}(\{\mathbf e_i:m_i\in S\})\right).
\end{aligned}
\end{equation}
Here, $\bar{\mathbf p}(S)$ is the Avg-mono reference. The mixture encoder applies a linear projection and molecule-level self-attention, followed by attention pooling with a learned query, an aggregation approach also used in AROMMA \cite{kang2026aromma}. 
The model adds a gated residual to the reference logits:

\begin{equation}
\label{eq:mixpom}
\begin{aligned}
\mathbf p_m(S)
&=\sigma\!\left(\operatorname{logit}\bar{\mathbf p}(S)
                 +\mathbf g_S\odot\Delta(\mathbf h_S)\right),\\
\mathbf g_S&=\tanh(\mathbf W_g\mathbf h_S).
\end{aligned}
\end{equation}
This formulation allows the model to increase or decrease descriptor probabilities according to the component combination. The residual head's final layer is initialized to zero, so training starts from Avg-mono.


BP annotations may omit odor attributes present in a mixture. We follow the class-distribution-aware pseudo-labeling strategy used in AROMMA \cite{kang2026aromma}, using frozen Avg-mono predictions to supplement training-pair labels. Class-specific thresholds are determined by descriptor frequencies in the single-molecule training set, while existing positive labels are retained. We train MixPOM with multi-label binary cross-entropy on the augmented targets and an additional consistency loss that keeps predictions close to Avg-mono on descriptors without direct BP supervision. Validation and test annotations remain unchanged. The detailed settings are provided in Appendix.

\subsection{Candidate Profile Construction}
\label{sec:indexing}

Given a molecular inventory, we precompute odor profiles for every single molecule and unordered pair to build an index for online retrieval.

For pairs, we use the five-fold mean MixPOM predictions on BP-supervised descriptors, with a bounded logit correction relative to Avg-mono to avoid extreme mixture corrections. Single-molecule candidates use Avg-mono predictions.

Retrieval uses the 138 OpenPOM descriptors. All candidate profiles are computed offline, so queries require no further molecular model inference.

\subsection{Language-Guided Matching}
\label{sec:matching}

Odor attributes in a query play different roles. For ``freshly baked strawberry cream cake,'' strawberry and cream define the main odor, sweetness provides background, and a fishy note clearly conflicts with the target. We therefore use a language model to map query $q$ to three pairwise disjoint descriptor sets:
\begin{itemize}
    \item {Lead}, $\mathcal D_l$: core attributes that define the target odor and receive the highest priority in retrieval;
    \item {Ambient}, $\mathcal D_a$: background attributes that support the overall odor impression and receive lower matching weights; and
    \item {Absent}, $\mathcal D_x$: attributes that clearly conflict with the target and incur a penalty.
\end{itemize}
All descriptors are drawn from a predefined vocabulary, and the lead set is nonempty. The three roles distinguish the main odor from its background and from clear conflicts. An attribute omitted from the query is not automatically considered undesirable: many odors include accompanying notes that the user may not mention. The complete prompt and output constraints are given in Appendix.

We compare each candidate profile $\mathbf p(S)$ with these requirements using a log-likelihood score. Desired attributes are scored by their predicted probabilities of being present, and absent attributes by their probabilities of being absent. We also penalize additional attributes according to how closely they are associated with the target odor.

\subsubsection{Likelihood of the Semantic Requirements}

Let $p_j(S)$ denote the predicted probability that candidate $S$ has attribute $j$. For lead and ambient attributes, we average the log-probabilities within each set:
\begin{equation}
\label{eq:positive_matching}
\begin{aligned}
R_l(S,q)&=\frac{1}{|\mathcal D_l|}\sum_{j\in\mathcal D_l}\log p_j(S),\\
R_a(S,q)&=\frac{1}{|\mathcal D_a|}\sum_{j\in\mathcal D_a}\log p_j(S).
\end{aligned}
\end{equation}
Taking the average prevents a role from receiving greater influence simply because it contains more descriptors. The logarithm makes low probabilities matter: for strawberry cream cake, a high strawberry probability alone does not ensure a good score if creamy is weakly supported. Increasing support for creamy would improve the score more than an equal increase in an already high strawberry probability. This encourages candidates to capture the different parts of the target. Ambient attributes are scored in the same way but receive a lower weight because they describe the background.

For an explicitly absent attribute, the required outcome has probability $1-p_j(S)$. We therefore use the mean log-probability of absence:
\begin{equation}
\label{eq:exclusion}
R_x(S,q)=\frac{1}{|\mathcal D_x|}\sum_{j\in\mathcal D_x}\log\big(1-p_j(S)\big).
\end{equation}
This term lowers the score when a candidate is predicted to contain a conflicting attribute. The penalty becomes stronger as that probability approaches one. Thus, a candidate with high strawberry and creamy probabilities can still receive a low score if fishy is also strongly predicted. As with the desired attributes, averaging controls the influence of the number of exclusions. An empty role contributes zero.

\subsubsection{Controlling Additional Attributes}
\label{sec:redundancy}

Additional attributes require a different treatment because they have not been explicitly excluded. Fruity, for example, may be a reasonable accompaniment to a strawberry target. Penalizing every unrequested attribute equally would discourage such combinations. We instead use descriptor co-activation to estimate which additional attributes are associated with the desired odor.

\begin{figure}[t]
    \centering
    \includegraphics[width=\columnwidth]{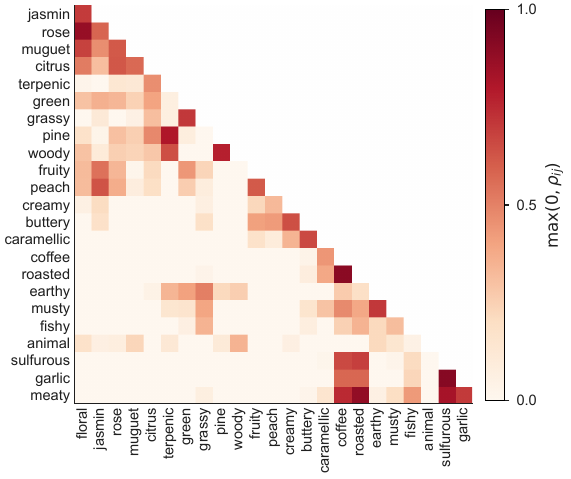}
    \caption{Descriptor co-activation for 24 selected attributes.}
    \label{fig:cooccurrence}
\end{figure}

We compute Pearson correlations between descriptor probabilities predicted by POM+ across GS-LF molecules and retain their nonnegative values. For each descriptor $j$, its relatedness to the query is the largest correlation with any lead or ambient descriptor:
\begin{equation}
\label{eq:coactivation}
\begin{aligned}
M_{jk}&=\max\!\left\{0,\operatorname{corr}_{m\in\mathrm{cache}}
\big(p_j(m),p_k(m)\big)\right\},\\
\rho_j(q)&=\max_{k\in\mathcal D_l\cup\mathcal D_a}M_{jk}.
\end{aligned}
\end{equation}
Using the largest correlation allows an additional note to be associated with any part of the target. It need not be related to every desired attribute to receive a smaller penalty. Negative correlations receive a value of zero and provide no reduction. Figure~\ref{fig:cooccurrence} shows a representative slice of the matrix. These associations affect only the treatment of additional attributes; the requested attributes are still scored using their own predicted probabilities.

Let $\mathcal D_o=\mathcal V\setminus(\mathcal D_l\cup\mathcal D_a\cup\mathcal D_x)$ be the unrequested descriptors in the scoring vocabulary $\mathcal V$. We penalize their mean predicted probability, weighted by their lack of association with the target:
\begin{equation}
\label{eq:redundancy}
R_o(S,q)=\frac{1}{|\mathcal D_o|}
\sum_{j\in\mathcal D_o}\big(1-\rho_j(q)\big)p_j(S).
\end{equation}
We set $R_o=0$ when $\mathcal D_o$ is empty. An attribute unrelated to the target contributes its full probability to the average, while a closely associated attribute contributes less. All unrequested descriptors are included, so the term reflects additional predictions across the candidate profile. Explicitly absent attributes are handled separately by $R_x$: their penalties are not reduced even if they correlate with a desired attribute.

\subsubsection{Final Ranking}

The final score combines the semantic log-likelihood terms with the additional-attribute regularizer:
\begin{equation}
\label{eq:score}
\begin{aligned}
r(S,q)={}&R_l(S,q)+\lambda_a R_a(S,q)\\
         &+\lambda_x R_x(S,q)-\lambda_o R_o(S,q).
\end{aligned}
\end{equation}
The lead coefficient is fixed to one. The weights $\lambda_a$, $\lambda_x$, and $\lambda_o$ balance background attributes, explicit conflicts, and additional notes. A candidate is favored when it supports the requested attributes, assigns low probabilities to clear conflicts, and limits unrelated additional attributes. Parameter values and numerical conventions are given in Appendix~\ref{app:scoring}.

At query time, the system parses the three semantic sets, scores every indexed candidate, and returns the highest-scoring single molecule or pair according to Equation~\ref{eq:retrieval}.


\section{Experiments}

We evaluate language-guided mixture retrieval, benchmark the odor predictors, and assess the contributions of MixPOM components through ablations. We also demonstrate physical preparation of formulations retrieved by olBlend.

\subsection{Experimental Setup}

\paragraph{Retrieval setting and baselines.}
We use the 89-odorant laboratory inventory and its 4,005 single-molecule and pair candidates. Across five open-ended queries, we compare olBlend with two baselines restricted to the same inventory.

{MixPOM-match} uses the hybrid candidate probabilities and the same frozen LLM lead set as olBlend. Let $\mathbf b_l\in\{0,1\}^{138}$ be the lead indicator, with $b_{l,j}=1$ exactly when $j\in\mathcal D_l$. The baseline selects
\begin{equation}
\label{eq:cosine_baseline}
S^*_{\mathrm{cos}}(q)=\arg\max_{S\in\mathcal C}
\frac{\mathbf p(S)^\top\mathbf b_l}
{\|\mathbf p(S)\|_2\|\mathbf b_l\|_2}.
\end{equation}
This cosine objective matches candidates to the lead indicator, with the predictor and lead interpretation held fixed across both methods.

{LLM-direct} selects one or two molecules directly from the query and inventory SMILES, without molecule names, odor labels, or predictor-based ranking. It provides a comparison with direct language-model selection. The complete parsing and selection prompts are given in Appendix.

\paragraph{Predictor data and evaluation.}
For single-molecule prediction, GS-LF contains about 5K molecules and 138 descriptors. We follow the official AROMMA five-fold protocol with a fixed test set of 1,007 molecules. POM and POM+ are trained from scratch on each fold's training molecules, excluding the test SMILES. For mixture prediction, each official fold contains approximately 3.0K single-molecule and 39.5K pair samples. POM+ is pretrained on about 5K odor-labeled GS-LF molecules and frozen for this mixture setting. MixPOM and AROMMA use the same official odor-supervised fold files; their architectures and encoder pretraining differ. BP evaluation uses the original human annotations, with no pseudo-label modification of test targets.

We report mean and standard deviation of macro AUROC across the five predictor folds, averaging over descriptors with both positive and negative test examples. Training aligns the two vocabularies to 152 dimensions; evaluation uses the respective task vocabulary. AUROC assesses descriptor discrimination, while retrieval also depends on the probability values and the semantic representation.

\subsection{Language-Guided Mixture Retrieval}
\label{sec:retrieval_results}

\begin{figure*}[t]
    \centering
    \includegraphics[width=0.98\textwidth]{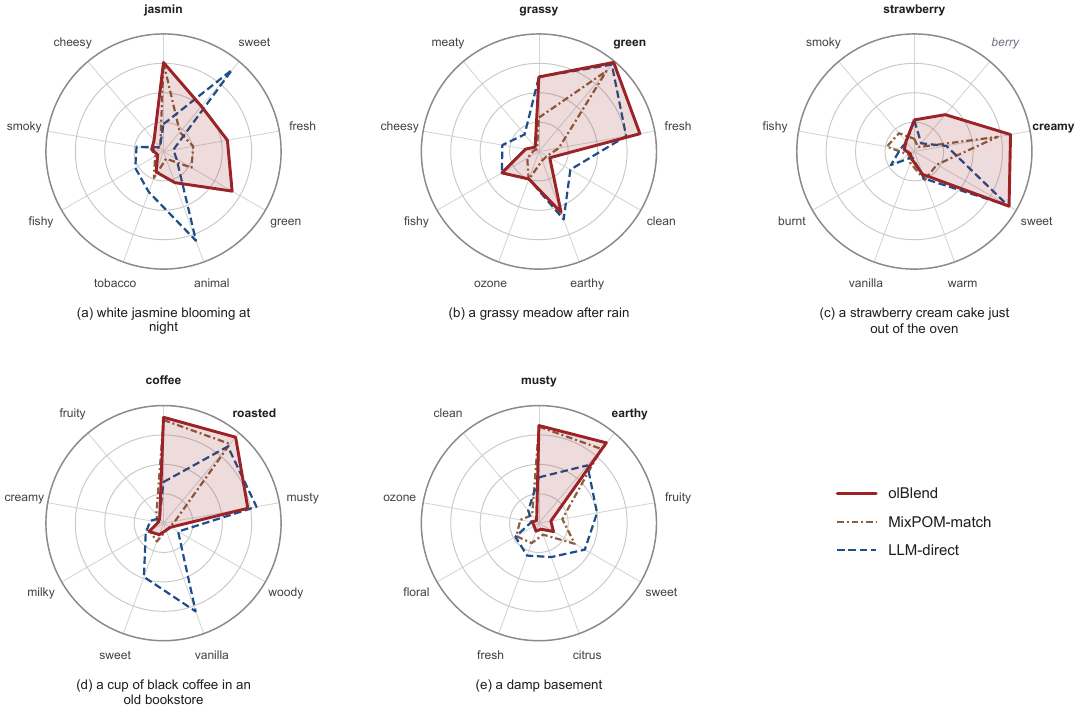}
    \caption{Model-based retrieval comparison across five queries.}
    \label{fig:query_radar}
\end{figure*}

Figure~\ref{fig:query_radar} compares the MixPOM-predicted profiles of the top-1 candidates selected by olBlend, MixPOM-match, and LLM-direct. The lead descriptors make the target identity explicit, allowing us to inspect whether the selected candidate supports that identity and what additional attributes it is predicted to carry.

Across the examples, olBlend generally selects profiles with higher probabilities for the lead descriptors, while several conflicting attributes are reduced relative to LLM-direct. MixPOM-match often produces similar lead probabilities to olBlend, with differences on supporting and additional descriptor axes. For example, the jasmine and meadow cases show differences on the green and fresh axes, respectively. These comparisons show that the structured objective brings background and exclusion requirements into candidate selection beyond lead-only matching.


\subsection{Evaluating the Odor Predictors}
\label{sec:prediction_results}

On GS-LF, POM+ improves macro AUROC from $0.864\pm0.002$ for POM to $0.877\pm0.001$ (Table~\ref{tab:mono_results}), providing a stronger single-molecule reference for mixture prediction.

\begin{table}[htbp]
\centering
\caption{Single-molecule odor prediction on GS-LF.}
\label{tab:mono_results}
\begin{tabular}{lc}
\toprule
Method & AUROC \\
\midrule
MPNN-GNN & $0.851 \pm 0.003$ \\
POM & $0.864 \pm 0.002$ \\
POM+ & $\mathbf{0.877 \pm 0.001}$ \\
\bottomrule
\end{tabular}
\end{table}

\begin{table}[htbp]
\centering
\caption{Mixture odor prediction on BP.}
\label{tab:mixture_results}
\begin{tabular}{lccc}
\toprule
Method & \# Single & \# Pair & AUROC \\
\midrule
MPNN-GNN
& 0
& $\sim$40K
& $0.734 \pm 0.019$ \\
Avg-mono
& $\sim$5K
& 0
& $0.828 \pm 0.007$ \\
AROMMA
& $\sim$50M
& $\sim$40K
& $0.861 \pm 0.006$ \\
\midrule
\textbf{MixPOM}
& $\sim$5K
& $\sim$40K
& $\mathbf{0.867 \pm 0.008}$ \\
\bottomrule
\end{tabular}
\end{table}

On BP, MixPOM reaches $0.867\pm0.008$ AUROC, compared with $0.828\pm0.007$ for Avg-mono (Table~\ref{tab:mixture_results}). This $0.039$ improvement shows that learning from mixture supervision improves descriptor prediction beyond averaging component predictions. For retrieval, the result motivates using mixture-specific profiles for pair candidates rather than relying entirely on their single-molecule references.

MixPOM is competitive with AROMMA ($0.861\pm0.006$) while using a single-molecule encoder trained on about 5K odor-labeled molecules rather than pretraining on about 50 million unlabeled molecules. 

\subsection{Mixture Predictor Ablations}
\label{sec:ablations}

Table~\ref{tab:mixpom_ablation} compares the full model with its ablations under the same five-fold protocol.
Removing the single-molecule prior skip while retaining the mixture encoder and learned prediction head gives $0.560\pm0.061$ AUROC. Setting the learned correction to zero instead recovers Avg-mono at $0.828\pm0.007$. Combining the prior with a learned mixture correction yields $0.867\pm0.008$, supporting the residual design of MixPOM.
\begin{table}[htbp]
\centering
\caption{MixPOM ablations on BP.}
\label{tab:mixpom_ablation}
\small
\begin{tabular}{lc}
\toprule
Model &  AUROC \\
\midrule
\textbf{MixPOM}
& $\mathbf{0.867 \pm 0.008}$ \\
w/o consistency regularization
& $0.866 \pm 0.012$ \\
w/o prior skip 
& $0.560 \pm 0.061$ \\
w/o tanh gate
& $0.858 \pm 0.005$ \\
w/o pseudo labels
& $0.844 \pm 0.012$ \\
Avg-mono
& $0.828 \pm 0.007$ \\
\bottomrule
\end{tabular}
\end{table}

Removing pseudo labels reduces AUROC from $0.867$ to $0.844$, while removing the tanh gate results in $0.858$. Removing consistency regularization gives $0.866$.

\subsection{Physically Prepared Samples}
\label{sec:prepared_samples}

\begin{figure}[H]
    \centering
    \includegraphics[width=\columnwidth]{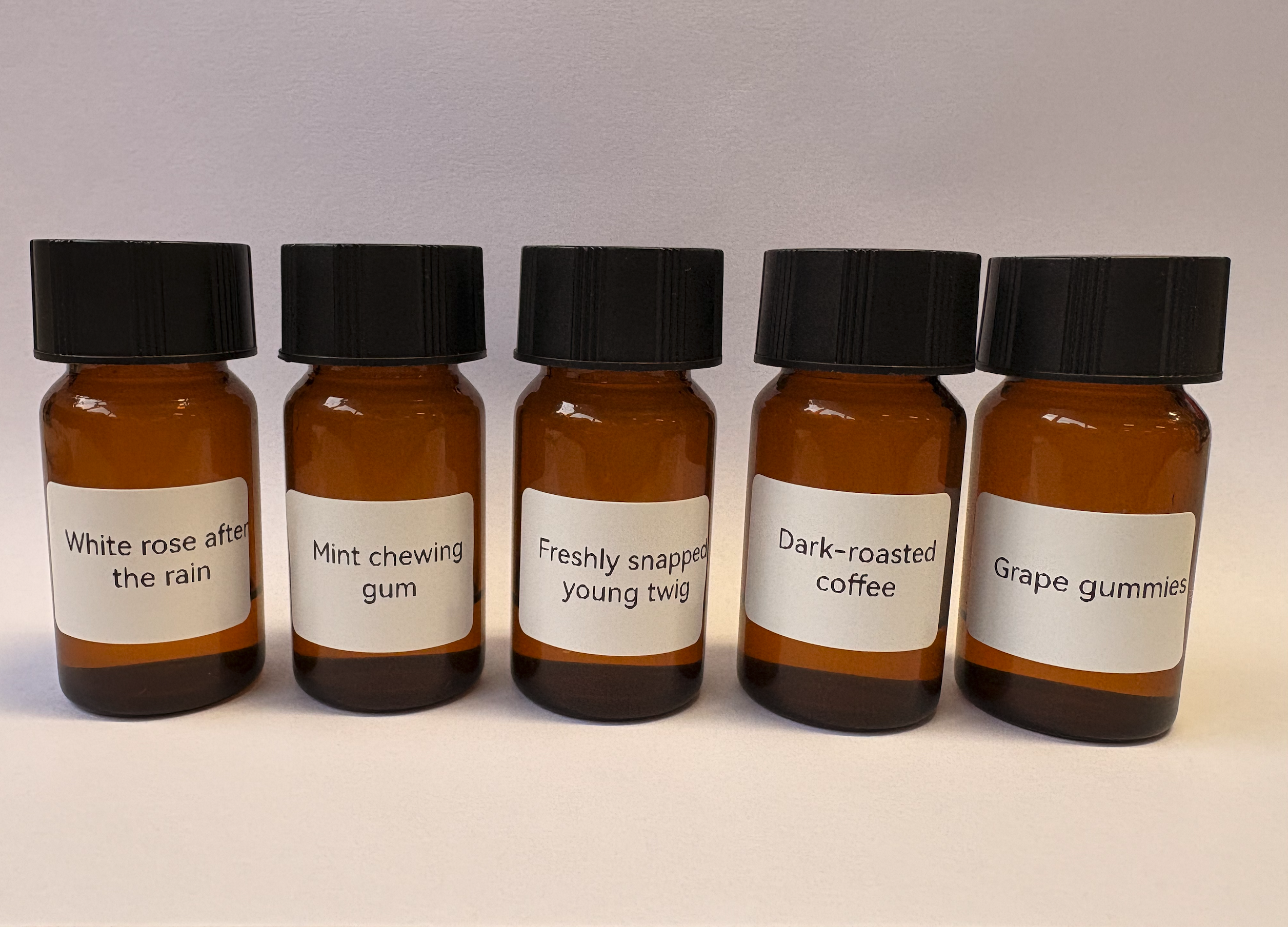}
    \caption{Physically prepared odor samples retrieved by olBlend (Table~\ref{tab:sensory_targets}).}
    \label{fig:sensory_samples}
\end{figure}

Five odor samples were physically prepared from the formulations retrieved by olBlend. Table~\ref{tab:sensory_targets} lists their target descriptions and sample identifiers, and Figure~\ref{fig:sensory_samples} shows the prepared samples.

\begin{table}[H]
    \centering
    \caption{Sample identifiers and target odor descriptions.}
    \label{tab:sensory_targets}
    \small
    \begin{tabular}{@{}lp{0.82\columnwidth}@{}}
        \toprule
        Sample & Target description \\
        \midrule
        S1 & White rose after the rain \\[3pt]
        S2 & Dark-roasted coffee with a slightly burnt, bitter note \\[3pt]
        S3 & A freshly opened bag of grape gummies \\[3pt]
        S4 & Freshly snapped young twig \\[3pt]
        S5 & Mint chewing gum \\
        \bottomrule
    \end{tabular}
\end{table}

\section{Conclusion}

olBlend translates natural-language odor requirements into molecular formulations drawn from an available inventory. POM+ and MixPOM supply candidate odor profiles, with MixPOM learning non-additive mixture corrections to a single-molecule prior. Structured retrieval jointly accounts for lead, ambient, and absent attributes, connecting these predictions to the user's intended odor. MixPOM improves BP AUROC from $0.828$ for Avg-mono to $0.867$. Retrieval across five open-ended queries over 4,005 candidates from 89 odorants and the physical preparation of five retrieved formulations demonstrate the workflow from language to molecular selection and sample production.

By separating language interpretation from molecular search, olBlend confines LLM processing to the odor query and performs retrieval through numerical scoring of precomputed candidate profiles. This design allows the molecular inventory to expand without enlarging the LLM input for each query. It provides a practical basis for language-guided formulation exploration and odor generation from available molecular resources.

\section{Limitations and Future Work}
Due to limitations of the available datasets, the current framework focuses on slightly over 100 odor descriptors. Some descriptors, however, may be inherently subjective and dependent on individual experience. For example, a descriptor such as ``sour'' can encompass a broad range of perceptual interpretations, and its olfactory association may vary across individuals. Future work could therefore explore more fine-grained and personalized representations of odor perception.

The current mixture setting follows the binary-mixture data available for training and evaluation \cite{sisson2025deep}, with concentration ratios held fixed. Future work will explicitly incorporate concentration and mixing ratios into mixture modeling and retrieval, and extend MixPOM and olBlend to ternary and higher-order formulations. As the inventory size and mixture order increase, we will also investigate more scalable search and generative strategies to address the combinatorial growth of candidate formulations.

\bibliographystyle{ieeetr}
\bibliography{references}

\clearpage

\appendix

\section{Hyperparameters and Implementation Settings}
\label{app:scoring}

\subsection{Hyperparameters}
Table~\ref{tab:hyperparameters} lists the settings for the model, training objective, and retrieval score. 
\begin{table}[H]
\centering
\small
\caption{Model, training, and retrieval settings.}
\label{tab:hyperparameters}
\begin{tabular}{cc}
\toprule
Parameter & Value \\
\midrule
Ensemble weight $\omega$ & $0.5$ \\
Embedding dimension & $256$ \\
Attention layers & $2$ \\
Consistency weight $\lambda_{\mathrm{cons}}$ & $0.5$ \\
Clipping scale $\beta$ & $1.5$ \\
Lead log-likelihood coefficient & $1.0$ \\
Ambient weight $\lambda_a$ & $0.3$ \\
Absent weight $\lambda_x$ & $1.0$ \\
Off-note weight $\lambda_o$ & $1.0$ \\
Logarithm floor $\epsilon$ & $10^{-8}$ \\
\bottomrule
\end{tabular}
\end{table}

\subsection{Vocabulary alignment}
Both POM+ branches produce 138-dimensional descriptor probabilities and 256-dimensional hidden states, which are averaged separately and cached. Let $\mathcal V_{\mathrm{mono}}$ and $\mathcal V_{\mathrm{BP}}$ denote the 138 GS-LF and 74 BP descriptors. Their union contains 152 descriptors. Single-molecule predictions are aligned by descriptor name: probabilities in $\mathcal V_{\mathrm{mono}}$ are copied directly, while the 14 BP-only descriptors receive a reference probability of $0.5$, corresponding to a zero logit. On these dimensions, the learned residual supplies the mixture prediction directly. The candidate index stores Avg-mono and the probability mean over five MixPOM folds. Retrieval uses the 138 OpenPOM descriptors. The matching implementation operates on the supplied scoring vocabulary and includes every descriptor outside the three semantic sets in the off-note regularizer; it does not separately exclude odorless.

\subsection{Retrieval details}
To limit the influence of large mixture corrections on retrieval scores, we apply a bounded correction in logit space. Let $\mathbf p_e(S)$ be the probability mean over the five MixPOM folds, $\bar{\boldsymbol\ell}=\operatorname{logit}\bar{\mathbf p}(S)$, and $\boldsymbol\ell_e=\operatorname{logit}\mathbf p_e(S)$. The retrieval logits are

\begin{equation}
\label{eq:retrieval_profile}
\ell_{r,j}
=\bar\ell_j+\beta\tanh\!\left(
\frac{\ell_{e,j}-\bar\ell_j}{\beta}\right).
\end{equation}
The scale $\beta=1.5$ bounds the logit correction. Pairs use $p_j(S)=\sigma(\ell_{r,j})$ on BP-supervised dimensions. Other dimensions and single-molecule candidates use $p_j(S)=\bar p_j(S)$.

\subsection{Pseudo-labeling and training objective}
Specifically, let $\pi_j$ be the positive-label frequency of descriptor $j$ in the current fold's GS-LF single-molecule training set. We choose a class-specific threshold $\tau_j$ so that approximately a fraction $\pi_j$ of training-pair predictions satisfy $\bar p_j(S)\geq\tau_j$, and augment the targets with positive pseudo labels:

\begin{equation}
\label{eq:pseudo_targets}
\widetilde y_j(S)=
\begin{cases}
1,&y_j(S)<0.5,\ \bar p_j(S)\geq\tau_j,\ \pi_j>0,\\
y_j(S),&\text{otherwise}.
\end{cases}
\end{equation}

For descriptors $\mathcal U$ without direct BP supervision, we additionally regularize mixture predictions toward Avg-mono to preserve the available single-molecule knowledge. The training objective combines multi-label binary cross-entropy with this consistency loss:

\begin{equation}
\label{eq:training_loss}
\begin{aligned}
\mathcal L_{\mathrm{cons}}
&=\frac{1}{|\mathcal U|}\sum_{j\in\mathcal U}
D_{\mathrm{KL}}\!\left[\mathrm{Bern}(\bar p_j)\parallel\mathrm{Bern}(p_{m,j})\right],\\
\mathcal L
&=\mathcal L_{\mathrm{BCE}}(\widetilde{\mathbf y},\mathbf p_m)
+\lambda_{\mathrm{cons}}\mathcal L_{\mathrm{cons}}.
\end{aligned}
\end{equation}

Here $\lambda_{\mathrm{cons}}=0.5$ controls the contribution of the consistency loss relative to the supervised binary cross-entropy term. The consistency loss averages the Bernoulli KL divergence from the frozen Avg-mono reference $\bar p_j$ to the MixPOM prediction $p_{m,j}$ over descriptors without direct BP supervision. A larger coefficient more strongly discourages deviations from this reference on those dimensions; setting it to zero removes consistency regularization, as in the corresponding ablation in Table~\ref{tab:mixpom_ablation}. This coefficient is used during predictor training and is distinct from the retrieval weights $\lambda_a$, $\lambda_x$, and $\lambda_o$ in Equation~\eqref{eq:score}.

BP-only descriptors have $\pi_j=0$ and receive no positive pseudo labels. The class frequency controls the proportion of teacher predictions selected, not the final positive-label rate after retaining original positives; it does not assume equal true prevalence in single molecules and mixtures. Unselected entries retain their stored targets. Single-molecule training targets and all validation and test targets are unchanged. Pseudo labels enter BCE through $\widetilde{\mathbf y}$ without a separate loss.

\section{LLM System Prompts}
\label{app:llm_prompts}

For reproducibility, this section reports the system prompts used for query parsing of olBlend and for the LLM-direct baseline. The LLM used is DeepSeek v4 flash.

\subsection{Prompt for olBlend}
\label{app:query_to_tags_prompt}



You are an odor-query understanding and descriptor-mapping system. Map a natural-language odor query to olfactory descriptors from the allowed vocabulary and return three ordered sets:

\begin{itemize}
    \item \texttt{lead}: the dominant identity of the target odor, containing one to three descriptors and receiving the highest retrieval weight;
    \item \texttt{ambient}: descriptors that are normally present in the scene but are contextual rather than identity-defining, and must not override the lead descriptors; and
    \item \texttt{absent}: descriptors that clearly conflict with the target odor.
\end{itemize}

\noindent Odor description: \texttt{\{description\}}

\noindent Allowed vocabulary: \texttt{\{vocab\}}

\noindent Follow these rules:

\begin{enumerate}
    \item \textbf{Understand the target odor.} The query may directly name an odor, material, object, place, or everyday scene. Infer what the user would typically expect to smell instead of relying only on keyword matching.

    \item \textbf{Construct \texttt{lead} (required).} The lead set answers: ``If only one to three words could be retained, what does this smell like?'' Include only dominant, perceptible, and distinctive odor identities. Use one to three descriptors, never more than three, ordered from most to least representative. Prefer specific odor sources or characteristic notes, such as \texttt{grassy}, \texttt{jasmin}, \texttt{coffee}, \texttt{strawberry}, \texttt{pine}, or \texttt{musty}, over generic adjectives. Do not use \texttt{clean}, \texttt{fresh}, \texttt{sweet}, \texttt{warm}, or \texttt{natural} alone as lead descriptors unless that attribute is itself the core of the query, as in ``clean linen.''

    \item \textbf{Construct \texttt{ambient} (optional).} Ambient descriptors are commonly present in the scene but should not determine the identity of the formulation, for example \texttt{fresh}, \texttt{clean}, \texttt{sweet}, \texttt{warm}, \texttt{earthy}, or \texttt{ozone}. They must not duplicate lead descriptors. Use zero to six ambient descriptors, and do not move a genuinely dominant odor into this set.

    \item \textbf{Construct \texttt{absent}.} Include a descriptor only when it clearly conflicts with the target. Use five to ten descriptors in typical cases, but return an empty set when there is no clear conflict. Do not exclude a descriptor merely because it is unpleasant.

    \item \textbf{Use direct evidence and controlled common sense.} A descriptor must be explicitly mentioned or be only one inferential step away from the named material or scene and dominant in everyday experience. The vocabulary describes odor rather than taste, texture, temperature, or humidity. Infer from those conditions only when they clearly indicate a perceptible odor source.

    \item \textbf{Map to the allowed vocabulary.} Use only the exact strings in the allowed vocabulary. Do not create, translate, paraphrase, combine, or inflect entries. Prefer an exact match; otherwise use the closest reasonable odor category. Ignore a cue when no reasonable match exists.

    \item \textbf{Resolve ambiguity.} When several interpretations are possible, choose the one most consistent with everyday experience and most likely to dominate actual olfactory perception. Prefer a smaller but more representative lead set.

    \item \textbf{Output constraints.} The \texttt{lead}, \texttt{ambient}, and \texttt{absent} sets must be pairwise disjoint, and \texttt{lead} must not be empty. Return only valid JSON in the following form:
\end{enumerate}

\begin{center}
\begin{minipage}{\columnwidth}
\centering\small
\texttt{\{"lead": ["descriptor\_1"],}\\
\texttt{"ambient": ["descriptor\_2"],}\\
\texttt{"absent": ["descriptor\_3"]\}}
\end{minipage}
\end{center}

\noindent Do not return Markdown, explanations, reasoning, confidence scores, or any other text.

\subsection{Prompt for LLM-Direct }
\label{app:query_to_formula_prompt}



You are a fragrance-formulation assistant. Given a natural-language odor query and a closed laboratory inventory, select the one or two molecules that best reproduce the requested odor.

\noindent Odor query: \texttt{\{description\}}

\noindent Inventory SMILES: \texttt{\{inventory\}}

\noindent Follow these rules:

\begin{enumerate}
    \item Return only exact SMILES strings that appear in the inventory. Do not invent, modify, translate, or substitute structures outside the inventory.

    \item The inventory provides structures only, without common names, IUPAC names, CAS numbers, or odor labels. Infer odor from chemical structure and do not assume names that were not provided.

    \item Select one molecule by default. Select two only when a single structure cannot cover the dominant odor identity of the query. Do not force a second molecule merely to make the formulation appear more complete.

    \item When two molecules are selected, treat the result as an unordered formulation and do not repeat the same SMILES.

    \item Do not return an empty list.
\end{enumerate}

\noindent Return only valid JSON in the following form:

\begin{center}
\begin{minipage}{\columnwidth}
\centering\small
\texttt{\{"smiles": ["SMILES\_1"],}\\
\texttt{"rationale": "one short sentence about the structures"\}}
\end{minipage}
\end{center}

\noindent Do not return Markdown, explanatory blocks, molecule names, or any other text.

\end{document}